\documentclass[prb,twocolumn,showpacs,graphics,preprintnumbers]{revtex4}
\usepackage{graphicx}
\usepackage{amsmath}
\usepackage{amssymb}
\def\cemin{Ce$M$In$_{5}$}
\def\ir{CeIrIn$_{5}$}
\def\rhir{CeRh$_{0.2}$Ir$_{0.8}$In$_{5}$}
\def\lair{LaIrIn$_{5}$}
\def\rh{CeRhIn$_{5}$}
\def\co{CeCoIn$_{5}$}

\begin{document}
\preprint{draft}

\title{Magneto-transport properties governed by the antiferromagnetic fluctuations in heavy fermion superconductor {\ir}}

\author{Y.~Nakajima,$^{1,2,\ast}$ H.~Shishido,$^{1}$ H.~Nakai,$^{1}$ T.~Shibauchi,$^{1}$ M.~Hedo,$^{2,\dag}$ Y.~Uwatoko,$^{2}$ T.~Matsumoto,$^{3}$ R.~Settai,$^{4}$ Y.~Onuki,$^{4}$ H.~Kontani,$^{5}$ and Y.~Matsuda$^{1,2}$}

\affiliation{$^1$Department of Physics, Kyoto University, Kyoto 606-8502, Japan }%
\affiliation{$^2$Institute for Solid State Physics, University of Tokyo, Kashiwanoha, Kashiwa, Chiba 277-8581, Japan}%
\affiliation{$^3$National Institute of Material Science, Sakura, Tsukuba, Ibaraki 305-0003, Japan}%
\affiliation{$^4$Graduate School of Science, Osaka University, Toyonaka, Osaka 560-0043, Japan}%
\affiliation{$^5$Department of Physics, Nagoya University, Furo-cho, Chikusa-ku, Nagoya 464-8602, Japan}%

\begin{abstract}

In quasi-two dimensional Ce(Ir,Rh)In$_5$ system, it has been suggested that the phase diagram contains two distinct domes with different  heavy fermion superconducting states.  We here report the systematic pressure dependence of the electron transport properties in the normal state of {\rhir} and {\ir}, which locates in first and second superconducting dome, respectively.   We observed non-Fermi liquid behavior at low temperatures in both compounds, including non-quadratic  $T-$dependence of the resistivity, large enhancement of the Hall coefficient, and the violation of the Kohler's rule in the magnetoresistance.   We show that the cotangent of Hall angle $\cot \Theta_H$ varies as $T^2$, and the magnetoresistance is quite well scaled by the Hall angle as $\Delta \rho_{xx}/\rho_{xx}\propto \tan^2\Theta_H$.   The observed transport anomalies are common features of {\cemin} ($M$=Co, Rh, and Ir) and high-$T_c$ cuprates, suggesting that the anomalous transport properties observed in {\ir} are mainly governed by the antiferromagnetic spin fluctuations, not by the Ce-valence fluctuations which has been proposed to be the possible origin for the second superconducting dome.

\end{abstract}


\maketitle

\section{Introduction} 

The resent discoveries of heavy fermion compounds {\cemin} ($M$=Rh, Co, and Ir) give a unique opportunity to elucidate the interplay between the magnetism and the superconductivity.  The ground state of these compounds can be tuned by pressure and chemical doping. {\co}\cite{petco} and {\ir}\cite{petir} are superconductors with the transition temperature $T_{c}=$ 2.3~K and 0.4~K at ambient pressure, respectively.  On the other hand, {\rh} is an antiferromagnet with $T_{N}=$ 3.8~K at ambient pressure and shows superconductivity under pressure.\cite{hegger} In {\co} and {\rh}, the thermodynamic and transport properties in the normal state  exhibit a striking deviation from conventional Fermi liquid behavior,\cite{sidorov,bianchi,tayama} which is commonly observed in the systems in the vicinity of the antiferromagnetic (AF) quantum critical point (QCP).  Then it is widely believed that the superconductivity  in {\rh} and {\co} is closely related to the AF fluctuations.

Recently, it has been suggested that CeIrIn$_5$ should be distinguished from CeCoIn$_5$ and CeRhIn$_5$, although all three compounds share similar quasi-two dimensional (2D) band structure.\cite{haga,settai}   Figure~1 depicts the schematic temperature -- $x$ ($T$-$x$)  phase diagram of CeRh$_{1-x}$Ir$_x$In$_5$ and  temperature -- pressure ($T$-$P$) phase diagram of CeIrIn$_5$.\cite{nicklas,kawasaki2}  In this system the Rh substitution for Ir increases the $c/a$ ratio, acting as a negative chemical pressure that increases AF correlations.  In CeRh$_{1-x}$Ir$_x$In$_5$, the ground state continuously evolves from AF metal  ($x<0.5$) to superconductivity ($x>0.5$). $T_c$ shows a maximum at $x\sim 0.7$ and exhibits a cusp-like minimum at $x \sim$ 0.9, forming a first dome (SC1).   The superconductivity nature in SC1, which occurs in the proximity to AF QCP, should be essentially the same as CeCo(In$_{1-x}$Cd$_x$)$_5$ \cite{pham} and CeRhIn$_5$.\cite{hegger}   The strong AF fluctuations associated with the AF QCP nearby are observed in SC1.\cite{kawasaki2,zhe01}  In CeIrIn$_5$ ($x$ = 1), $T_c$ increases with pressure and exhibits a maximum ($T_c$ = 1~K) at $P\sim$ 3~GPa, forming a second dome (SC2).  The AF fluctuations in SC2 far from the AF QCP are strongly suppressed, compared with those in SC1.\cite{kawasaki2,zhe01,kawasaki1}  Moreover, it has been reported that the nature of the crossover behavior from non-Fermi to Fermi liquid in strong magnetic fields for CeIrIn$_5$ is very different from that for CeCoIn$_5$ and CeRhIn$_5$.\cite{cap04,pag03,par06}

From the analogy to CeCu$_{2}$(Si$_{1-x}$Ge$_{x}$)$_{2}$ with two distinct superconducting domes,\cite{yuan} a possibility that the Ce-valence fluctuations play an important role for the normal and superconducting properties in {\ir} has been pointed out.\cite{watanebe}  For instance, it has been suggested that while the superconductivity in SC1 is magnetically mediated,  the superconductivity in SC2 may be mediated by the Ce-valence fluctuations.\cite{holmes2}   Thus the major outstanding question is whether the Ce-valence fluctuations play an important role for the physical properties of {\ir} in SC2 phase.   Our previous studies indicate that the transport coefficients, including resistivity, Hall effect, and magnetoresistance, can be powerful tools to probe the AF spin fluctuations.\cite{nakajima1,nakajima2,nakajima3}  In this paper, we report the systematic pressure study of the transport properties for {\rhir} and {\ir}, which locates in SC1 and SC2 phase, respectively.    We provide several pieces of evidence that all the anomalous transport properties observed in {\ir} and {\rhir} originate from the AF spin fluctuations irrespective of which superconducting phase the system belongs to.

\begin{figure}[tb]
\begin{center}
\includegraphics[width=8cm]{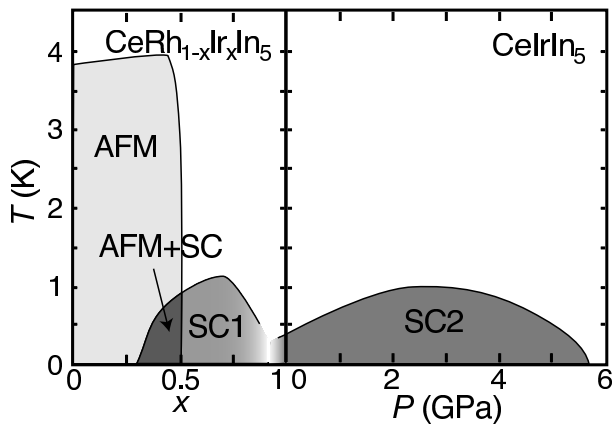}
\end{center}
\caption{Schematic $T$-$x$ phase diagram for CeRh$_{1-x}$Ir$_{x}$In$_{5}$ and $T$-$P$ phase diagram for {\ir} \cite{nicklas,kawasaki2,chemp}  .}
\label{PD}
\end{figure}

\section{Experimental}

\begin{figure}[b]
\begin{center}
\includegraphics[width=8cm]{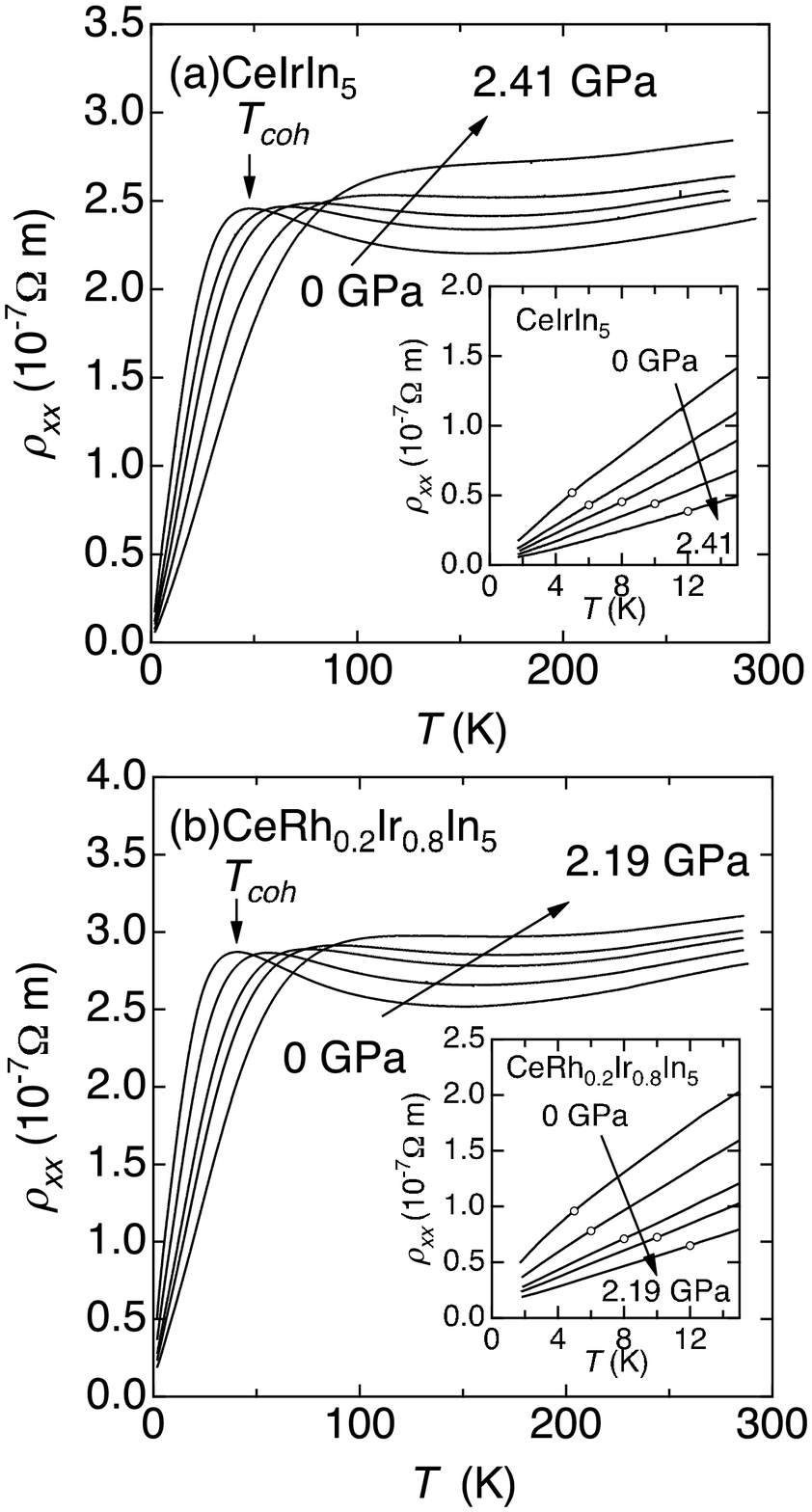}
\end{center}
\caption{(a) Temperature dependence of resistivity for {\ir} at 0, 0.56, 0.98, 1.59, and 2.41GPa.  (b) Temperature dependence of resistivity for {\rhir} at 0, 0.49, 1.11, 1.50, and 2.19GPa.  Insets are expanded views at low temperatures. Downarrows drawn in main panels indicate $T_{coh}$ at ambient pressure, where the resistivity shows a broad maximum. Open circles shown in the insets indicate the resistivity at $T_{m}$ where $R_{H}$ shows a minimum. For detail, see the text in \S.4.}
\label{rhoxxT}
\end{figure}

The high quality single crystals of {\ir} and {\rhir} were grown by the self-flux method.  We performed all measurements on samples with a typical dimension of $\sim1.0\times$2.0$\times$0.1 mm$^{3}$ in the transverse geometry for {\boldmath $H$} $\parallel c$ and the current {\boldmath $J$} $\parallel a$.  The Hall effect and transverse magnetoresistance were measured simultaneously.  We obtained Hall resistivity from the transverse resistance by subtracting the positive and negative magnetic field data.  Hydrostatic pressure up to 2.41~GPa were generated in a piston-cylinder type high pressure cell with oil as a transmitting fluid (Daphne 7373 : petroleum ether = 1 : 1). The pressure inside the cell was determined by the superconducting transition temperature of Pb.

\section{Results}
\subsection{Resistivity}

Figures \ref{rhoxxT}(a) and (b) show the temperature dependence of the resistivity $\rho_{xx}$ in zero field at several pressures for {\ir} and {\rhir}, respectively.  The overall feature of the temperature dependence for both compounds is typical in Ce-based heavy fermion compounds.  On cooling from room temperature, $\rho_{xx}$ first decreases and then increases due to dominant Kondo scattering.  At lower temperatures, $\rho_{xx}$ exhibits a metallic behavior after showing a broad maximum at around the temperature $T_{coh}$, shown by arrows.  $T_{coh}$ corresponds to the Fermi temperature of $f$ electrons and the system becomes coherent below $T_{coh}$.   $T_{coh}$ increases with pressure.  The insets of Figs. \ref{rhoxxT}(a) and (b) show the low temperature data of $\rho_{xx}$ for {\ir} and {\rhir}, respectively.   The resistivities of {\ir} and {\rhir} are markedly different from the $T^{2}$-behavior expected in Fermi liquid metals.  At ambient pressure, $\rho_{xx}$ varies as 
\begin{equation}
\rho_{xx}\sim T^{\alpha}
\label{eq:rhoxx}
\end{equation}
with $\alpha\sim 1$ for {\ir} and $\sim 0.7$ for {\rhir}.  $\alpha$ increases with pressure for both systems and reaches  $\sim 1.4$ at 2.4~GPa for {\ir} and $\sim 1.3$ at 2.19~GPa for {\rhir}, indicating that the Fermi liquid behavior is recovering by applying pressure.    We note that these $\alpha$-value is close to that reported in Ref.~\onlinecite{muramatsu}.   These temperature and pressure dependence of resistivities for {\ir} and {\rhir} are very similar to those of {\co} and {\rh}.\cite{nakajima2, nakajima3}

\subsection{Hall effect}

\begin{figure}[b]
\begin{center}
\includegraphics[width=8cm]{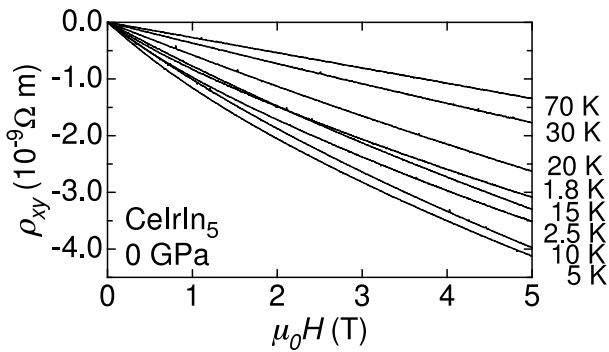}
\end{center}
\caption{Field dependence of $\rho_{xy}$ for {\ir} at ambient pressure.}
\label{rhoxyH}
\end{figure}

\begin{figure}[t]
\begin{center}
\includegraphics[width=7cm]{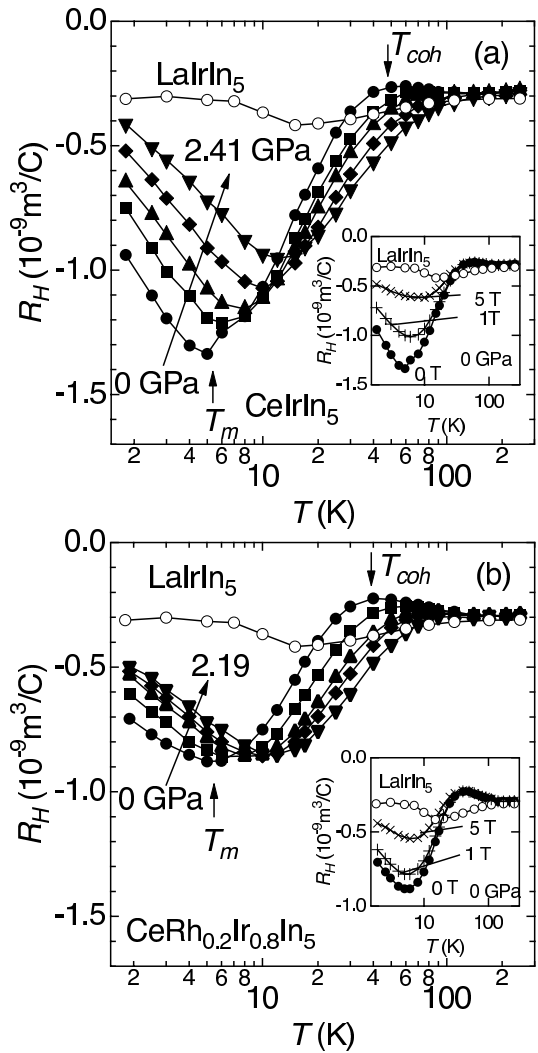}
\end{center} 
\caption{(a) Temperature dependence of $R_{H}$ for {\ir} at several pressures (0 ($\bullet$), 0.56  ($\blacksquare$), 0.98 ($\blacktriangle$), 1.56 ($\blacklozenge$), and 2.41 GPa ($\blacktriangledown$)) and for {\lair} at ambient pressure ($\circ$). $R_{H}$ is defined by the zero-field limit for derivative of $\rho_{xy}$. Inset: Temperature dependence of $R_{H}$ for {\ir} at 0 ($\bullet$), 1 ($+$), and 5 T ($\times$) at ambient pressure. (b)Temperature dependence of $R_{H}$ for {\rhir} at several pressures (0 ($\bullet$), 0.49  ($\blacksquare$), 1.11  ($\blacktriangle$), 1.50  ($\blacklozenge$) and 2.19 GPa ($\blacktriangledown$)) and for {\lair} at ambient pressure ($\circ$). Inset: Temperature dependence of $R_{H}$ for {\rhir} at 0 ($\bullet$), 1 ($+$), and 5 T ($\times$) at ambient pressure. Down and up arrows in main panels indicate $T_{coh}$ at ambient pressure determined by the resistivity peak and $T_{m}$ at ambient pressure, where $R_{H}$ shows a minimum, respectively.}
\label{RHT}
\end{figure}

Figures \ref{rhoxyH} depict the Hall resistivity $\rho_{xy}$ as a function of magnetic field at ambient pressure for {\ir}.  The sign of $\rho_{xy}$ is negative.  At low temperatures, $\rho_{xy}$ deviates from the $H$-linear dependence. Similar behavior is observed in {\rhir}.  Figures \ref{RHT}(a) and (b) show the temperature dependence of the Hall coefficient  $R_{H}$ in zero field limit defined as $R_{H}\equiv\lim_{H\rightarrow 0}\frac{d\rho_{xy}}{dH}$  at several pressures for {\ir} and {\rhir}, respectively.   For comparison,  $R_{H}$ of {\lair}, which has no $f$-electron and has similar band structure to {\ir}, is  plotted in the same figure.  For {\lair}, $R_H$ shows a shallow minimum at around 20~K and becomes nearly $T$-independent at low temperatures.  The carrier number estimated from $R_{H} \sim 3 \times 10^{-10}$m$^{3}$/C for {\lair} at low temperatures corresponds to nearly three electrons per unit cell, which is consistent with the number expected from the band structure, indicating $R_H\simeq 1/ne$ where $n$ is the carrier number.

The Hall effect in {\ir} and {\rhir} is distinctly different from that in {\lair}.  The temperature dependence of $R_{H}$ for {\ir} and {\rhir} is closely correlated with the resistivity.  The down-arrow in Figs. \ref{RHT} (a) and (b) indicates $T_{coh}$ at ambient pressure determined by the resistivity peak in Figs.~2(a) and (b), respectively.  In the high temperature regime above $T_{coh}$, $R_{H}$ for {\ir} and {\rhir} shows weak $T$-dependence.  Well above $T_{coh}$, $R_H$ for both compounds well coincides with $R_{H}$ of {\lair}, indicating $R_H \simeq 1/ne$.  Below $T_{coh}$,  $R_{H}$ for {\ir} and {\rhir} decreases rapidly with decreasing $T$.   At lower temperatures,  $R_H$ increases after showing minimum at $T_{m}$ indicated by up-arrows in Figs.~\ref{RHT}(a) and (b).  With increasing pressure, $T_m$ increases and the enhancement of $|R_{H}|$ at low temperature regime is reduced for {\ir}.

\begin{figure}[t]
\begin{center}
\includegraphics[width=7cm]{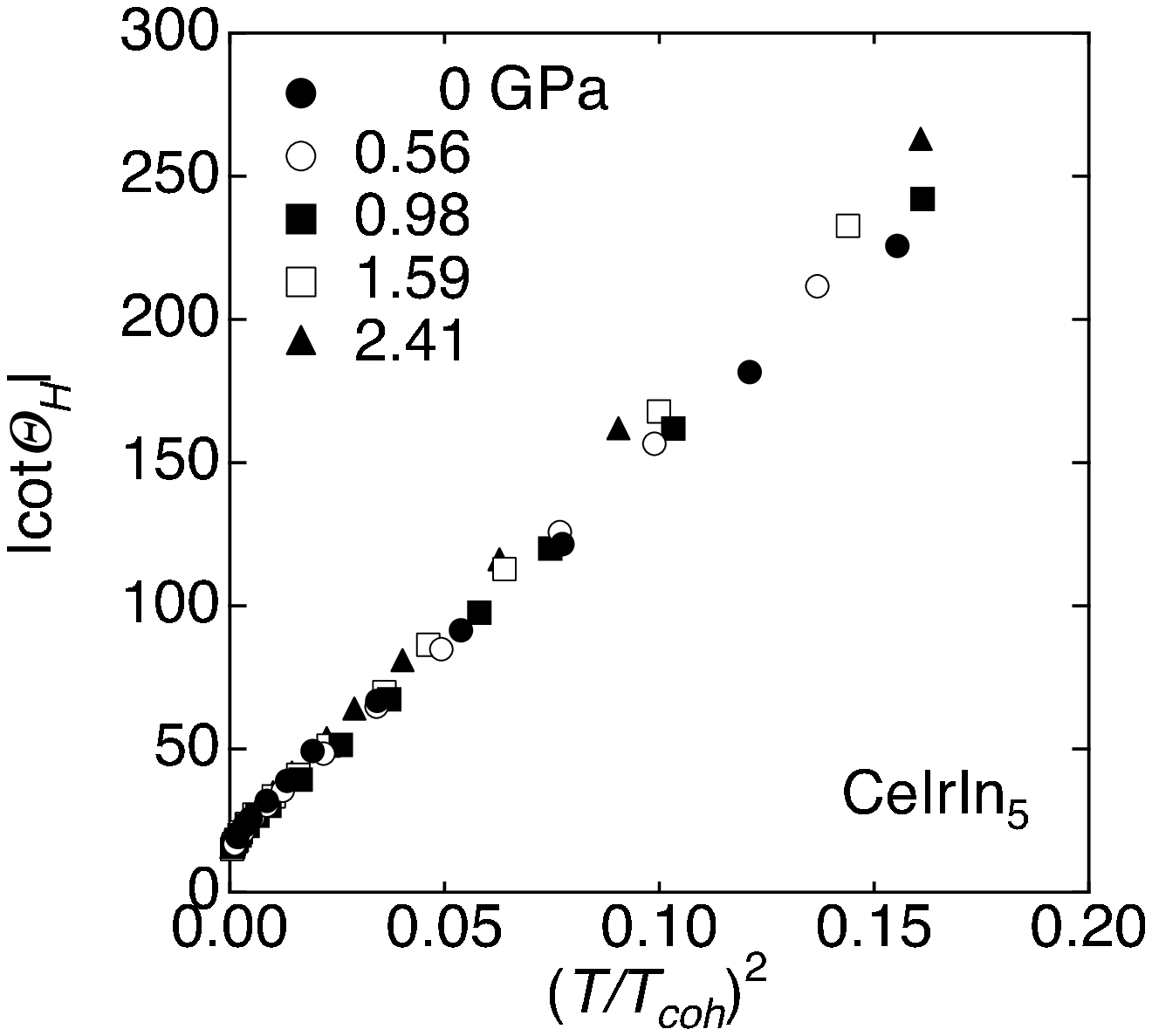}
\end{center} 
\caption{$|\cot\Theta_{H}|$ as a function of $(T/T_{coh})^{2}$ for {\ir} at 0 ($\bullet$), 0.56 ($\circ$), 0.98 ($\blacksquare$), 1.59 ($\square$), and 2.41 GPa ($\blacktriangle$).}
\label{cot}
\end{figure}

The insets of Figs.~\ref{RHT} (a) and (b) show the temperature dependence of $R_{H}$ at $\mu_0H$=0, 1, and 5 T at ambient pressure for {\ir} and {\rhir}, respectively.  $R_{H}$ is defined by a field derivative of $\rho_{xy}$, $R_{H}\equiv d\rho_{xy}/dH$.  The magnitude of $R_{H}$ below $T_{coh}$ is strongly suppressed by magnetic fields.

We here comment on the effect of the skew scattering.  Usually, $R_{H}$ in heavy fermion compounds can be written by the sum of the ordinary Hall part $R_{H}^{n}$ due to Lorentz force and the anomalous Hall part $R_{H}^{a}$ due to skew scattering,\cite{fert} 
\begin{equation}
	R_{H}=R_{H}^{n}+R_{H}^{a}.
\end{equation}	
The magnitude of $R_H^a$ is often much larger than that of $R_H^n$ except for $T\ll T_{coh}$ and $T\gg T_{coh}$.   In most Ce-based heavy fermion systems, $R_H^a$ is positive in sign and shows a strong $T$-dependence, which is scaled by $\chi\rho_{xx}$ (Ref.~\onlinecite{fert}) or $\chi$.\cite{konyam}  At around $T_{coh}$,  $R_{H}^a$ shows a broad maximum and its amplitude becomes much larger than $1/|ne|$.  It it obvious that $R_{H}$ of {\ir} and {\rhir} are very different from that expected from the skew scattering.   In fact, the sign of $R_{H}$ is negative in the whole temperature regime.   Moreover, $R_{H}$ is close to $1/ne$ at $T\sim T_{coh}$.  A slight increase of $R_H$ observed at $T\gtrsim T_{coh}$ in the low pressure regime appears to come from small but finite contribution of the skew scattering. Thus the skew scattering contribution is small in {\ir} and {\rhir} and the normal part of Hall effect is dominant.  We also note that skew scattering is negligibly small in {\rh} and {\co}.\cite{nakajima1,nakajima2,nakajima3}

The Hall effect in {\ir} and {\rhir} below $T_{coh}$, particularly the enhancement of $|R_H|$ from $|1/ne|$, is distinctly different from that expected in the conventional metals.  Such an enhancement has also been reported in {\co} and {\rh}, and high-$T_c$ cuprates.    There, it has been shown that the Hall problem can be simplified when analyzed in terms of Hall angle $\Theta_H\equiv \tan^{-1}\frac{\rho_{xy}}{\rho_{xx}} $; $\cot \Theta_H$ well obeys a $T^2$-dependence,
\begin{equation}
\cot \Theta_H=AT^2+B,
\end{equation}
where $A$ and $B$ are constants.\cite{chien,anderson,hussey}  We here examine $\cot \Theta_H$ for {\ir}.   Figure \ref{cot} depicts $\cot \Theta_H$ as a function of $T^2$ for {\ir}. In the all pressure regime,  $\cot \Theta_H$ well obeys a $T^2$-dependence, except for the low temperature regime, exhibiting a striking similarity with {\co} and {\rh}, and  high-$T_c$ cuprates.

\subsection{Magnetoresistance}

\begin{figure}[b]
\begin{center}
\includegraphics[width=8cm]{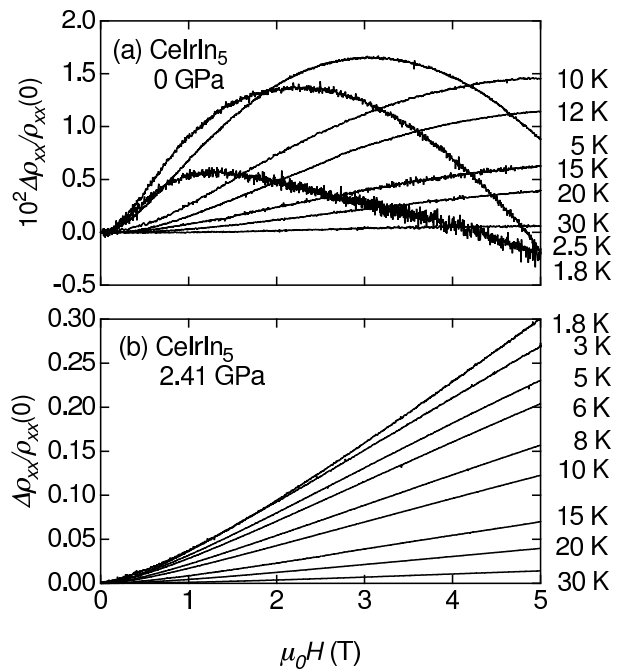}
\end{center}
\caption{Magnetoresistance of {\ir} as a function of $H$ at (a) 0 GPa and (b) 2.41 GPa.}
\label{rhoxxH}
\end{figure}

Figures \ref{rhoxxH} (a) and (b) show the magnetoresistance $\Delta\rho_{xx}/\rho_{xx}(0)\equiv(\rho_{xx}(H)-\rho_{xx}(0))/\rho_{xx}(0)$ of {\ir} at ambient pressure and at $P$ = 2.41~GPa, respectively.   The magnetoresistance varies as $\Delta\rho_{xx}/\rho_{xx}(0)\propto H^{2}$ at very low field ($\mu_0H<0.2$~T).  At ambient pressure, the magnetoresistance decreases with $H$ at high fields below 5~K.  This phenomena has also been observed in {\co} at ambient pressure\cite{nakajima1, nakajima3} and is attributed to the spin-flop scattering.

\begin{figure}[t]
\begin{center}
\includegraphics[width=8cm]{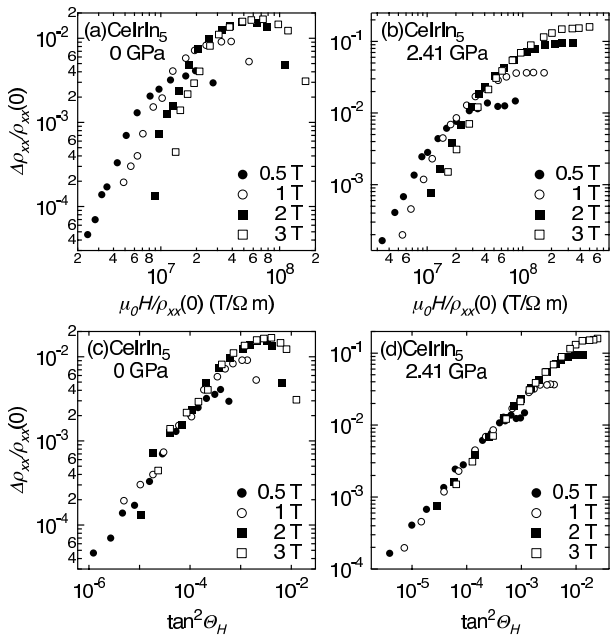}
\end{center}
\caption{Kohler's plot. $\Delta\rho_{xx}/\rho_{xx}(0)$ vs $\mu_{0}H/\rho_{xx}(0)$ for {\ir} at (a)0GPa, and (b)2.41GPa. Modified Kohler's plot. $\Delta\rho_{xx}/\rho_{xx}(0)$ as a function of $\tan^{2}\Theta_{H}$ for {\ir} at (c)0GPa and (d)2.41GPa.}
\label{MR}
\end{figure}

We here discuss the magnetoresistance in the conventional and unconventional metals.  In conventional metals, the magnetoresistance due to orbital motion of carriers  obeys the Kohler's rule,
\begin{equation}
	\frac{\Delta\rho_{xx}(H)}{\rho_{xx}(0)}=F\left(\frac{\mu_{0}H}{\rho_{xx}(0)}\right),
\end{equation}
where $F(x)$ is a function which depends on the details of electronic structure.\cite{pippard}  It has been shown that the magnetoresistance in LaRhIn$_5$ with similar electronic structure but with weak electronic correlation well obeys the Kohler's rule.\cite{nakajima3}   We first test the validity of the Kohler's rule in the magnetoresistance of {\ir}.   Figures \ref{MR}(a) and (b) depict $\Delta\rho_{xx}/\rho_{xx}(0)$ of {\ir} as a function of $\mu_{0}H/\rho_{xx}(0)$ at 0 and 2.41 GPa, respectively.   The data never collapse into the same curve, indicating a  violation of the Kohler's rule.

A striking violation of the Kohler's rule has been reported in {\co} and {\rh}, and  high-$T_c$ cuprates.  It has been shown instead that in these systems the magnetoresistance is well scaled by $\tan^{2}\Theta_{H}$ (modified Kohler's rule), where $\Theta_{H}\equiv\tan^{-1}(\rho_{xy}/\rho_{xx})$ is the Hall angle;\cite{nakajima1, nakajima3,harris}
\begin{equation}
	\frac{\Delta\rho_{xx}}{\rho_{xx}(0)}\propto \tan^{2}\Theta_{H}.
\end{equation}
We then examine the validity of the modified Kohler's rule for {\ir}.  In Figs.~\ref{MR}(c) and (d), the same data of magnetoresistance are plotted as a function of $\tan^{2}\Theta_{H}$.  For both cases, the data collapse into the same curve for three orders of magnitude, indicating that the magnetoresistance well obeys modified Kohler's rule.  The deviation from the modified Kohler's rule is observed at low temperature and high field region, possibly due to the spin-flop scattering.

\section{Discussion}

Summarizing the salient features in the transport properties of {\ir} below $T_{coh}$, which corresponds to the Fermi temperature of $f$ electrons,

\begin{enumerate}
\item[(i)] The dc-resistivity shows non-quadratic dependence, $\rho_{xx}\propto T^{\alpha}$ with $\alpha$ close to unity at ambient pressure.
\item[(ii)]  $|R_H|$  increases with decreasing temperature and reaches a value much larger than $|1/ne|$ well below $T_{coh}$.    The  Hall angle  varies as $\cot\Theta_H\propto T^2$.   
\item[(iii)] Magnetoresistance displays $T$- and $H$- dependence that strongly violates the Kohler's rule, 	$\Delta\rho_{xx}(H)/\rho_{xx}(0)\neq F(\mu_{0}H/\rho_{xx}(0))$.  Magnetoresistance well obeys the modified Kohler's rule that indicates a  scaling by the  the Hall angle, ${\Delta\rho_{xx}}/{\rho_{xx}} \propto \tan^2\Theta_H$.

\end{enumerate}
It should be emphasized that all of these features bear striking resemblance to those observed in {\co}, {\rh}, and high-$T_c$ cuprates.  Therefore, it is natural to consider that the transport properties commonly observed in {\ir} originate from the same mechanism.

Our previous studies have shown that the anomalous features in the transport phenomena (i)--(iii)  can be accounted for in terms of the recent theory in which the anisotropic inelastic scattering due to AF spin fluctuations are taken into account.   In the presence of strong AF fluctuations, the transport scattering rate strongly depends on the position of the Fermi surface.  Then the hot spots, at which the electron scattering rate is strongly enhanced by the AF fluctuations, appear at the positions where the AF Brilouin zone intersects with the Fermi surface.    The presence of the hot spots has been confirmed in high-$T_c$ cuprates and CeIn$_3$.\cite{ebihara}   Since the hot spot area does not contribute to the electron transport, it reduces the effective carrier density, which results in the enhancement of the $|R_H|$ from $|1/ne|$.  In such a situation, various transport properties are determined by $\tau_{cold}$, where $\tau_{cold}$ is the scattring time of the cold spots on the Fermi surface,  at which the electrons are less scattered.   Moreover, it has been shown that the transport properties are modified by the {\it backflow} accompanied with the anisotropic inelastic scattering.\cite{kontani,kanki,kontaniMR,ROP}

According to Refs.~\onlinecite{kontani,kanki,kontaniMR,ROP}, the transport properties under magnetic fields are governed by the AF correlation length $\xi_{AF}$ in the presence of backflow effect.   Zero-field diagonal conductivity $\sigma_{xx}(0)$, Hall conductivity $\sigma_{xy}$ and magnetoconductivity $\Delta\sigma_{xx}(H)\equiv \sigma_{xx}(H)-\sigma_{xx}(0)$ are given as
\begin{equation}
	\sigma_{xx}(0)\sim \tau_{cold},
\end{equation}
\begin{equation}
	\sigma_{xy} \sim \xi_{AF}^2\tau_{cold}^2H, \label{eq:sigxy}
\end{equation}
and
\begin{equation}
	\Delta\sigma_{xx}\sim \xi_{AF}^4\tau_{cold}^3H^{2}.\label{eq:sigxx}
\end{equation}
Here, we have dropped the higher terms with respect to $\tau_{cold}H$ since 
$\Delta\rho/\rho_0\ll1$ in the present experiment, which suggests that 
the relation $\omega_c\tau\ll1$ is satisfied. In the presence of AF fluctuation,  $\xi_{AF}$ depends on $T$ as $\xi_{AF}^2 \propto 1/(T+\theta)$, where $\theta$ is the Weiss temperature. Moreover, according to AF spin fluctuation theory, $\tau_{cold}$ is nearly inversely proportional to $T$; $\tau_{cold}\propto 1/T$.\cite{kontani}   When $T\gg\theta$, we then obtain the temperature dependence of the resistivity, $\rho_{xx}=\sigma_{xx}^{-1}$, Hall coefficient, $R_{H}=\sigma_{xy}/\sigma_{xx}^2H$,  and the Hall angle as
\begin{equation}
	\rho_{xx} \propto\tau_{cold}^{-1} \propto T,
\label{eq:rho}
\end{equation}
\begin{equation}
	R_{H} \propto \xi_{AF}^2 \propto \frac{1}{T},
\label{eq:Rh}
\end{equation}
and 
\begin{equation}
	\cot\Theta_H \propto T^2.
\label{eq:cot}
\end{equation}
By definition, the magnetoresistance is given by
\begin{equation}
	\frac{\Delta\rho_{xx}(H)}{\rho_{xx}(0)}=- \frac{\Delta\sigma_{xx}(H)}{\sigma_{xx}(0)}-\left (\frac{\sigma_{xy}(H)}{\sigma_{xx}(0)}\right )^2.   
\end{equation}
Using the relation $\Delta\sigma_{xx}(H)/\sigma_{xy}(H)^2\sim \tau_{cold}^{-1}$, given by Eqs.~(\ref{eq:sigxy}) and (\ref{eq:sigxx}), the magnetoresistance is obtained as
\begin{equation}
	\frac{\Delta\rho_{xx}(H)}{\rho_{xx}(0)}=(\tan\Theta_H)^2\cdot\left(\frac{\sigma_{xx}(H)}{\sigma_{xx}(0)}\right)^{2}\cdot(C-1),
\label{eq:mag}
\end{equation}
where $C$ is a constant and is $\sim$ 10-100 for Ce$M$In$_5$.   Since $\sigma_{xx}(H)/\sigma_{xx}(0)\simeq 1$ at low fields, $\Delta\rho_{xx}(H)/\rho_{xx}(0)$ is well scaled by $\tan^2 \Theta_H$.  Thus Eqs. (\ref{eq:rho}), (\ref{eq:Rh}), (\ref{eq:cot}), and (\ref{eq:mag}) reproduce the salient features of resistivity, Hall coefficient, Hall angle, and magnetoresistance observed in {\ir}, respectively.

The $H$-dependence of $R_H$ shown in the insets of Fig.~\ref{RHT}(a) reinforces the conclusion that the AF fluctuations govern the electron transport phenomena in {\ir} (also in {\rhir}).   The enhancement of $|R_H|$ below $T_{coh}$  is strongly suppressed  by magnetic fields and approaches that of $R_H$ of LaIrIn$_5$. This is consistent with the recovery of the Fermi liquid state in magnetic fields in {\ir}. Similar phenomena have also been reported in {\co} and {\rh}, where the Fermi liquid state is recovered in magnetic fields by the suppression of AF fluctuations.\cite{nakajima3}

The upturn behavior of $R_{H}$ for {\ir} and {\rhir} at low temperatures below $T_{m}$ shown in Figs.~\ref{RHT} (a) and (b) is also observed in {\co}.\cite{nakajima1,nakajima2,nakajima3}  This phenomenon can be explained by the reduction of {\it backflow} effect due to the effect of the  impurity scattering.   Below $T_{m}$,  isotropic impurity scattering becomes dominant and the {\it backflow} effect due to anisotropic scattering is relatively reduced.\cite{nakajima2,nakajima3}  To obtain more insight into the impurity effect, we compare the resistivity values at $T_m$.  The small open circles in the insets of Figs.~\ref{rhoxxT} (a) and (b) indicate the resistivity at $T_{m}$ where $R_{H}$ shows a minimum.  The values of the resistivity at $T_m$ are nearly pressure independent and close to $\sim 5~\mu\Omega$cm and $\sim 8~\mu\Omega$cm in {\ir} and {\rhir}, respectively.  We note that these values are close to the values of $\rho_{xx}$ at $T_{m}$ for {\co}.

We here discuss the difference between {\ir} and CeCu$_{2}$(Si$_{1-x}$Ge$_{x}$)$_{2}$.  For CeCu$_{2}$(Si$_{1-x}$Ge$_{x}$)$_{2}$  in the second superconducting dome,  anomalous behavior in transport and thermodynamic properties are observed near the pressure $P_{v}$, where $T_{c}$ shows a maximum.   For instance,  $\alpha$ in Eq. (\ref{eq:rhoxx}) approaches unity and  residual resistivity $\rho_{0}$ exhibits a maximum near $P_{v}$.\cite{yuan}   For {\ir}, on the other hand, $\alpha$ approaches the Fermi liquid value at $P\sim$ 3 GPa, where $T_{c}$ shows a maximum.  Moreover, the residual resistivity decreases with pressure as shown in the insets of Fig. \ref{rhoxxT}, which could be caused by the backflow or enhancement of impurity scattering near AF QCP.\cite{ROP}  These results indicate that there seems to be crucial differences in the transport phenomena between {\ir} and  CeCu$_{2}$(Si$_{1-x}$Ge$_{x}$)$_{2}$.

The presence of the AF fluctuations in {\ir} has been reported by the measurements of the nuclear magnetic resonance (NMR) relaxation rate $T_1^{-1}$.  According to NMR results,  AF fluctuations are strongly suppressed with pressure and there is no indication of the AF fluctuations at $P\gtrsim$ 1 GPa.\cite{kawasaki2}  The present results indicate that the transport measurements are more sensitive to the AF fluctuations than NMR experiments.

We finally comment on the superconducting gap structure in {\ir}.  Recent measurements of the anisotropy of the inter- and in-plane thermal conductivity for {\ir} suggest a hybrid gap structure,\cite{shakeripour} whose symmetry is different from $d_{x^2-y^2}$ for {\co} (Refs.~\onlinecite{izawa,mat06,green}) and (most probably) for {\rh}.  However,  very recent thermal conductivity measurements under rotated magnetic fields suggest that the superconducting gap structure for {\ir} has $d_{x^{2}-y^{2}}$ symmetry,\cite{kasahara} which implies that the AF spin fluctuations are important for the occurence of the superconductivity for {\ir}, which is consistent with the present work.

\section{conclusion}

We have investigated the detailed electron transport properties by applying pressure in the normal state of {\rhir} and {\ir}, which locates in the first and second superconducting dome, respectively.   We observed striking non-Fermi liquid behaviors below $T_{coh}$, including non-quadratic  $T-$dependence of the resistivity, large enhancement of the Hall coefficient at low temperatures $(|R_H|\gg1/|ne|)$, and the violation of the Kohler's rule in the magnetroresistance $\Delta\rho_{xx}(H)/\rho_{xx}\neq F(\mu_0H/\rho_{xx})$.    Moreover, we showed that the cotangent of Hall angle $\cot \Theta_H$ varies as $T^2$, and the magnetoresistance is quite well scaled by the Hall angle as $\Delta \rho_{xx}/\rho_{xx}\propto \tan^2\Theta_H$.  These non-Fermi liquid properties, particularly the Hall effect, are suppressed by pressure and magnetic fields.   The observed transport anomalies are common features of {\ir}, {\co}, {\rh}, and high-$T_c$ cuprates.  These results lead us to conclude that the non-Fermi liquid behavior observed in the transport properties in {\ir} originates not from the Ce-valence fluctuations but from the low-lying excitation due to the  AF fluctuations that still remain in the second dome away from the first dome in the proximity to the AF QCP.

\section*{Acknowledgment}
We thank H.~Ikeda, K.~Miyake and S.~Watanabe for stimulating discussions.  This work was partly supported by a Grant-in-Aid for Scientific Reserch from the Ministry of Education, Culture, Sports, Science and Technology.

\end{document}